\documentclass{optica-article}
\DeclareUnicodeCharacter{2212}{-}

\usepackage[utf8]{inputenc}

\journal{opticajournal} 

\articletype{Research Article}

\usepackage{lineno}

\usepackage{amssymb}
\usepackage{amsmath}
\usepackage{subcaption}
\usepackage{siunitx}
\usepackage{newunicodechar}
\newunicodechar{μ}{\mu}

\begin{document}

\title{High-reflectivity homoepitaxial distributed Bragg reflectors for photonic applications}

\author{Helena Janowska,\authormark{*} Anna Musiał, and Grzegorz Sęk}

\address{Laboratory for Optical Spectroscopy of Nanostructures, Department of Experimental Physics, Faculty of Fundamental Problems of Technology, Wrocław University of Science and Technology, Wybrzeże Wyspiańskiego 27, 50-370 Wrocław, Poland}

\email{\authormark{*}268275@student.pwr.edu.pl}


\begin{abstract*} 
Distributed Bragg reflectors (DBRs) are one of the basic photonic structures used to define microcavities for fundamental light-matter coupling studies, as well as to optimize performance of optoelectronic and photonic devices, e.g., lasers or non-classical light sources. The reflectivity of these structures depends critically on the refractive index contrast between the two quarter-wavelength thick layers constituting the DBR. At the same time, epitaxial fabrication process limits the choice of materials to those with the same, or very similar lattice constant to avoid strain accumulation in the relatively thick multilayer structure. This becomes very often a bottleneck for the DBR designs at certain wavelengths or for some of the material systems. Therefore, we explore theoretically DBR designs employing the reflective index contrast between undoped and doped layers of the same material, making the entire growth process homoepitaxial. The refractive index for a doped layer is calculated taking into account the free carrier absorption, carrier-carrier interaction, the Burnstein-Moss and plasma effects. The reflectivity spectrum of a DBR is further calculated using transfer matrix method. Exemplary results for three application relevant materials - hBN, InP and Si suitable for different spectral ranges, i.e. ultraviolet, telecommunication and mid-infrared, respectively, are presented. We report reflectivities on the level of 90\% for technologically achievable doping concentrations and moderate number of layer pairs.   

\end{abstract*}

\section{Introduction}
Distributed Bragg reflectors (DBRs) are well-known as essential components in both fundamental research and advanced photonic applications. 
They enable key studies of cavity quantum electrodynamics effects, polariton physics~\cite{Schneider2013}, and Bose-Einstein condensates formation~\cite{Pieczarka2024}, and are used in both, monolithic and open cavity~\cite{Tomm2021} designs. 
High-quality DBR-based resonators are used in non-classical light sources to improve the emission directionality, photon indistinguishability, and extraction efficiency~\cite{Rickert2025}.
They are also integrated into optoelectronic devices like micro- and nanolasers~\cite{Kadhim2025}, quantum photonic chips and integrated circuits~\cite{Burla2013}.
Typically, they are made in a single epitaxial growth process.
The structure contains alternating layers, usually of two different materials, with distinct refractive indices.
Through the strategic selection of materials and their respective thicknesses, precise control over the central wavelength of the high reflectivity stopband, can be achieved. 
However, there are certain limitations to the materials of choice: they should be as much as possible lattice-matched (also to the substrate), exhibit high refractive index contrast and no absorption in the spectral range of the gain medium~\cite{Paschotta}.
These conditions cannot always be easily met, and hence satisfying DBR performance can be difficult to achieve for specific spectral range or material system. For example, for InP substrate epitaxy, used, for instance, for growth of quantum dots emitting in the 3rd telecommunication window, reflectivity exceeding 90\% can only be achieved with quaternary alloy with specific composition requiring the growth control with ultra-high precision~\cite{Benyoucef2013,Wyborski2021}. 

In this paper, we propose DBR structures comprising layers of one material with different carrier concentrations. The refractive index, and consequently the reflectivity spectra, rely on the carrier concentration within the material as a result of three primary phenomena: bandfilling, bandgap renormalization, and free-carrier absorption.
This characteristic can be altered, in particular, through doping, interaction with external fields, or carrier injection~\cite{Bennett1990}.
We explore this fact as the basis for novel DBR designs with the benefit of homoepitaxial growth where the lattice matching condition is fulfilled automatically.
Our materials of choice are hexagonal boron nitride, indium phosphide and silicon, all application-relevant as elaborated below.

Hexagonal boron nitride (hBN) belongs to two-dimensional materials similar to graphene, with equal amounts of boron and nitrogen atoms organized in a hexagonal lattice. 
It forms a layered structure that is held together by weak van der Waals forces. 
The wide indirect band gap of $E_g = 5.955$~eV~\cite{Cassabois2016} makes it extensively used in transition metal dichalcogenide (TMDC) technology, as it acts as an encapsulating layer, insulating layer or tunnel barrier~\cite{Britnell2012}. 
On the other hand, hBN has its potential in the field of photonics, as it may be used in optoelectronic devices operating in the deep ultraviolet spectral range~\cite{Jiang2014}. 
Lately, it has also been proven that defects in hBN layers can act as single-photon sources in various spectral ranges~\cite{Tran2015}, which shows its potential in quantum communication technologies. 
All-hBN DBR structures were first realized by Else, A. Ciesielski et al.~\cite{Ciesielski2024} who obtained refractive index contrast by different porosity levels in adjacent layers. Their approach resulted in structures consisting of 15.5 layer pairs with a reflectivity that reaches 87\%.

Indium phosphide (InP) is a III-V semiconductor material that crystallizes in a zinc-blende structure.
Its direct band gap of $E_g=1.334$~eV and high electron mobility of $\mu = 5400$~$cm^2V^{-1}s^{-1}$ make it highly suitable for electronic and optoelectronic applications, particularly in devices operating in the near-infrared spectral range. 
Due to these properties, InP is widely used in the fabrication of high-speed photodetectors~\cite{Beling2009}, lasers~\cite{Sprengel2013}, and integrated photonic circuits~\cite{Smit2019}.
Moreover, InAs/InP quantum dots have attracted significant interest as efficient single-photon emitters capable of generating indistinguishable photons~\cite{Holewa2020,Holewa2024} or entangled photon pairs~\cite{Salter2010}, which is essential for quantum communication networks and quantum information processing.
Additionally, InP is well suited for heterogeneous integration with silicon platform, enabling the development of scalable quantum photonic integrated circuits (QPICs)~\cite{Li2025,Holewa2022}.
All these features position InP as a key material in the development of novel optoelectronic devices and components for quantum communication and cryptography, especially in the telecom spectral range.
The idea of DBR comprising layers of the same material with different doping levels has recently been proposed by Badura et. al~\cite{Badura2024}. They grew homoepitaxial InP-based structures reflecting around 99.5\% in the MIR spectral range. To obtain such high reflectivity at rather low photon energy and hence in a range of increased free-carrier absorption, they optimized the thicknesses of the layers to maintain the constructive interference condition for a pair of layers and the doping gradient to maximize the refractive index contrast.

Silicon (Si) is a foundation material for today's electronic systems due to its availability, well-developed processing techniques, and excellent electrical properties.
It is a semiconductor with an indirect energy gap of $E_g=1.12$~eV and electron mobility around $1400$~$cm^2V^{-1}s^{-1}$.
The indirect bandgap limits its efficiency in light emission, making its application in optoelectronic devices challenging.
Regardless of this, Si-based photonic circuits, devices, and platforms have been reported before~\cite{Joyce2024}. As the goal nowadays is to develop light sources compatible with Si either by heterogeneous bonding~\cite{Holewa2022} or by direct growth~\cite{Rudno2024, Viavmitinov2020,Nanwani2025} of efficient emitters on Si or in Si.
Regarding DBRs in Si-based structures, one can mention $SiO_2/TiO_2$ Bragg mirrors.
Feng et al.~\cite{Feng2013} fabricated and characterize these types of DBR, obtaining 95\% of reflectivity at 1.55~$\mu m$.

In the first section, we will describe the methodology of our calculations.
It contains a step-by-step procedure for estimating the change in the absorption coefficient and refractive index due to different free carrier concentrations in a material.
We also mention calculating reflectance of a layered structure using the transfer matrix method (TMM) and show an approach to include absorption.
In the next section, we present results on absorption and refractive index change for different carrier concentrations, as well as reflectivity spectra of designed DBRs for selected emitters compatible with proposed DBR material.
The selected light sources include hBN, which emits around 230~nm~\cite{You2020}, as well as InAs/InP and InAs/AlGaInAs/InP quantum dots, emitting near 1300~nm and 1550~nm, respectively.
Emission in the UVC spectral range is particularly valuable due to its well-known disinfection and antibacterial properties. It is also utilized as a component of white light emitters, as well as in sensing technologies and fluorescence-based analytical systems.~\cite{Hsu2021}.
Emitters operating at telecommunication wavelengths — typically around 1300~nm and 1550~nm — are critical for enabling low-loss, high-speed data transmission through optical fibers. These wavelengths correspond to the low-attenuation windows of silica fibers, making them ideal for long-distance optical classical as well as quantum communication.
Moreover, they are well suited for integration with silicon-based photonic technologies, as this spectral range lies within the transparency windows of Si, $SiO_2$, and $SiN_x$ - materials commonly used for fabricating passive components in integrated photonic circuits.
Additionally, we are exploring the mid-infrared region around 8000 nm, where strong absorption features of harmful gases such as $CO_2$, $NH_3$, and $CH_4$ occur~\cite{Lim2024, Munsaka2024}.
The spectral range around 8 $\mu m$ is of interest also for its relevance to free-space transmission window.
Recent studies bring the concept of non-classical light sources within the MIR domain, where DBRs will play a crucial part in maximizing photon extraction efficiency, in particular via microcavity designs ~\cite{Iles-Smith2025}.
Significance of various effects affecting the refractive index for different characteristic spectral ranges is also described.
The following part covers a discussion, including a comparison to DBRs of conventional designs.

\section{Method}
\begin{figure}[htbp]
\centering\includegraphics[width=7cm]{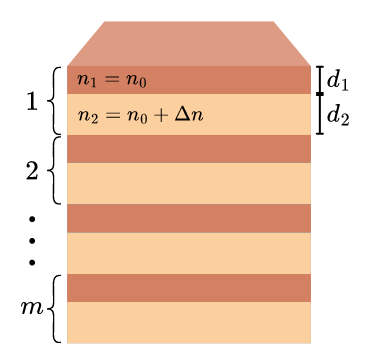}
\caption{Schematic of a distributed Bragg reflector structure.}
\label{DBR_sructure}
\end{figure}

In our work, we propose and explore theoretically a distributed Bragg reflector structure, which scheme is shown in Fig.~\ref{DBR_sructure}. 
We are considering two types of layers of the same material. 
The first has a refractive index $n_0$ corresponding to undoped material, and the second has a refractive index $n_0+\Delta n$, where $\Delta n$ is the difference in refractive index caused by the introduction of additional free carriers into the valence or conduction band. 
In general, it can be performed either by carrier injection or doping. 
In the case of doping, the carrier concentration in the corresponding band (typically lower than the dopant concentration) is relevant for the refractive index change in our calculations.
The undoped and doped layers have thicknesses of $d_1$ and $d_2$, respectively. 
To optimize the reflectance, the number of layer pairs $m$ is also varied.

To evaluate the influence of carrier concentration on the refractive index, we calculate $\Delta n$ by including three main effects that appear when doping a material. First, one has to calculate the influence of additional carriers on absorption to translate into $\Delta n$ using Kramers-Kronig relations. 
The first effect is the bandfilling or Burstein-Moss effect. 
Excess carriers are filling the bottom of the conduction or valence band, resulting in a shift of Fermi level and increased effective/optical band gap energy~\cite{Bennett1990}. 
This effect causes an increase in the refractive index for photon energies below the absorption edge and a decrease for energies above the absorption edge. 
The change of absorption coefficient related to this effect can be expressed by:
\begin{equation}\label{eq0}
\begin{split}
\Delta \alpha_{bf}=\frac{C}{E}\sqrt{E-E_g}[f_v(E_{a})-f_c(E_{b})-1]
\end{split}
\end{equation}
for direct bandgap semiconductors, and:
\begin{equation}\label{eq1}
\begin{split}
\Delta \alpha_{bf}=\frac{C_{1}}{E}(E-E_g+E_{ph})^2[f_v(E_{a})-f_c(E_{b})-1] \\
+\frac{C_{2}}{E}(E-E_g-E_{ph})^2[f_v(E_{a})-f_c(E_{b})-1]
\end{split}
\end{equation}
for indirect ones.
E is the energy of the incident photon, $E_g$ is bandgap of undoped material, $E_{ph}$ is phonon energy, $f_c$ and $f_v$ are probabilities of conduction and valence band being occupied by an electron at energies $E_b$ and $E_a$, respectively.
$C$, or $C_1$ and $C_2$ are fitting parameters of the absorption coefficient spectra of intrinsic (undoped) material. 
For direct bandgap material, the absorption coefficient is described by formula:
\begin{equation}\label{eq1.5}
\alpha = \frac{C}{E}\sqrt{E-E_g}
\end{equation}
and for indirect:
\begin{equation}\label{eq2}
\alpha = \frac{C_1}{E}(E-E_g+E_{ph})^2+\frac{C_2}{E}(E-E_g-E_{ph})^2
\end{equation}
Two segments of equation represent photon absorption with the participation of emitted and absorbed phonon, respectively.

Secondly, we include the bandgap renormalization effect. 
The high carrier concentration results in an increase in the Coulomb interactions within respective bands - between holes in the valence band and electrons in the conduction band. 
Repulsive interactions between carriers cause screening and induce shrinking of the band gap.
As a result, we obtain the redshift in the bandgap energy. 
The effect is followed by a corresponding change in absorption coefficient:
\begin{equation}\label{eq2.5}
\begin{split}
    \Delta \alpha_{br}=\frac{C}{E}\sqrt{E-E_g+\Delta E_g(X))}-\frac{C}{E}\sqrt{(E-E_g)}
\end{split}
\end{equation}
for direct bandgap materials, and for indirect:
\begin{equation}\label{eq3}
\begin{split}
    \Delta \alpha_{br}=\frac{C_1}{E}(E-E_g+E_{ph}+\Delta E_g(X))^2-\frac{C_1}{E}(E-E_g+E_{ph})^2\\
    +\frac{C_2}{E}(E-E_g-E_{ph}+\Delta E_g(X))^2-\frac{C_2}{E}(E-E_g-E_{ph})^2
\end{split}
\end{equation}
where $\Delta E_g$ is the energy difference between the bandgap energy of undoped and doped material caused by band gap renormalization, which depends on carrier concentration. 
The renormalization of bandgap $\Delta E_g$ for non-degenerate semiconductors (starting point of our considerations) is expressed by~\cite{Mahata2015}:
\begin{equation}\label{eq4}
    \Delta E_g(X)=\frac{3e^3}{16\pi\epsilon}\sqrt{\frac{X}{\epsilon kT}}
\end{equation}
where e is electron charge, T is temperature, $k=8.617\cdot 10^{5}$~eV/K is Boltzmann constant, $\epsilon$ is permittivity and X is free carriers' concentration - electrons or holes.
Effectively, the bandgap renormalization has the opposite effect on refractive index in comparison to bandfilling.
Near the bandgap energy, the change in the refractive index depends on the mutual ratio of these two phenomena.

We also include the plasma effect, which corresponds to intra-band free carrier absorption at low energies - smaller than the bandgap energy. 
To calculate the refractive index difference caused by this effect, we adopted the Drude model:
\begin{equation}\label{eq5}
    \Delta \alpha_{fca}=\frac{e^3\lambda^2X}{4\pi^2c^3\epsilon_0n_0m^{*2}\mu}
\end{equation}
where $\lambda$ is wavelength of incident photon, $m^*$ is effective mass, $\mu$ is carrier mobility, $\epsilon_0$ is permittivity of vacuum and $c$ is speed of light~\cite{Schroder1978}.
Materials constants used in this work are showed in Table~\ref{tab:materials}.
In the first row, references to the absorption spectra used to determine the C coefficients are listed.
For hBN and InP, the experimental data were obtained from conventional reflectivity and transmission measurements~\cite{Zunger1976,Turner1964}.
In contrast, the absorption coefficients for Si were derived from refractive index data obtained via ellipsometry~\cite{Green1995}.

\smallskip

\begin{table}[htbp]
\centering
\caption{\bf Materials constants of hBN, InP and Si.}
\begin{tabular}{cccc}
\hline
Material parameters & hBN & InP & Si \\
\hline
$\alpha_0(cm^{-1})$ spectra$^a$ & \cite{Zunger1976} & \cite{Turner1964} & \cite{Green1995} \\
$C(eV^{-1/2}cm^{-1})$ & - & $1.2548\cdot 10^5$ & - \\
$C_1(eV^{-1}cm^{-1})$ & $5.9838\cdot 10^9$ & - & $8.6603\cdot 10^3$ \\
$C_2(eV^{-1}cm^{-1})$ & $1.642\cdot 10^7$ & - & $6.9725\cdot 10^3$ \\
$E_g(eV)$ & 5.955~\cite{Cassabois2016} & 1.344~\cite{Ioffe} & 1.12~\cite{Ioffe} \\
$E_{ph}(eV)$ & 0.17~\cite{Hoffman1984} & - & 0.0641~\cite{Ioffe}\\
$\epsilon_r$ & 6.9~\cite{Laturia2018} & 12.5~\cite{Ioffe} & 11.7~\cite{Ioffe} \\
$m_e^*$ & 0.26~\cite{Ioffe} & 0.08~\cite{Ioffe} & 0.98~\cite{Ioffe} \\
$m_{hh}^*$ & - & 0.6~\cite{Ioffe} & 0.49~\cite{Ioffe} \\
$m_{lh}^*$ & - & 0.09~\cite{Ioffe} & 0.16~\cite{Ioffe} \\
$m_h^*$ & 0.47~\cite{Ioffe} & - & - \\
$n_0$ & 2.1~\cite{Grudinin2023} & 3.1~\cite{Adachi1989} & 3.4~\cite{Franta2018} \\
$\mu(cm^2/(V\cdot s)$ & 444~\cite{Khatami2021} & 5400~\cite{Ioffe} & 1400~\cite{Ioffe} \\
$N_C(cm^{-3})$ & $2.1\cdot 10^{19}$~\cite{Levinshtein2001} & $5.7\cdot 10^{17}$~\cite{Ioffe} & $3.2\cdot 10^{19}$~\cite{Ioffe} \\
$N_V(cm^{-3})$ & $2.1\cdot 10^{19}$~\cite{Levinshtein2001} & $1.1\cdot 10^{19}$~\cite{Ioffe} & $1.8\cdot 10^{19}$~\cite{Ioffe} \\
\hline
\end{tabular}
  \label{tab:materials}\\
$^a$Absorption coefficient spectra of intrinsic material used to determine $C$ or $C_1$ and $C_2$ coefficients.
\end{table}

To obtain a change in the refractive index, we use the Kramers-Kronig (K-K) relation, which transforms the imaginary part of the complex refractive index function to a real one. 
The total $\Delta n$ is a sum of refractive index changes caused by three effects discussed above:
\begin{equation}\label{eq6}
    \Delta n=\Delta n_{bf}+\Delta n_{br} +\Delta n_{fca}
\end{equation}
We performed calculations in a self-written Matlab program, where K-K relation was implemented with a script from Lucarini's book on data analysis~\cite{Lucarini2024}.

To optimize the DBR performance for a target spectral range, one has to calculate the reflectivity spectra for given parameters: carrier concentration related to refractive index contrast, thicknesses of the two layers, and number of DBR layer pairs. 
We use a Transfer Matrix Method implemented in Python. 
This method involves considering the electric field amplitudes during the propagation of an electromagnetic wave through the structure. 
The propagation within materials is characterized by specific refractive indices and thicknesses, as well as reflection and transmission at each interface. 
These are described by 2 x 2 matrices of coefficients transforming the amplitudes of electric field on one side of the structure to the amplitudes of electric field on the opposite side of the structure. 
The propagation of light throughout an entire structure is expressed by a two-dimensional array called a transfer matrix, which is calculated by multiplying all 2 x 2 matrices. 
The reflectivity is given by:
\begin{equation}\label{eq7}
    R=(\frac{M_{21}}{M_{11}})^2
\end{equation}
where $M_{11}$ and $M_{12}$ represents incident and reflected light~\cite{Muriel1997}.

We expanded the standard TMM approach to include absorption, which is a direct consequence of including plasma effect and therefore the possibility of absorption of photons with energies below the absorption edge due to intraband absorption. 
In the first approximation it is included by applying the Lambert-Beer law.
As a result, the final expression describing reflectivity at a specific wavelength is:
\begin{equation}\label{eq8}
    R=(\frac{M_{21}}{M_{11}})^2exp(-(\alpha(n_1)d_1+\alpha(n_2)d_2+...))
\end{equation}
where $\alpha$ is a total absorption coefficient and d is a layer thickness.
Effectively, the plasma effect $\Delta \alpha_{fca}$ is the only one having an impact on $\alpha$ in the proposed DBR structures.
The subscripts of $n$ and $d$ refer to the characteristics of the specific layers shown in Fig.~\ref{DBR_sructure}.
All calculations were performed assuming room temperature conditions (300~K).

\section{Results} 
\begin{figure}[htbp]
\centering\includegraphics[width=13cm]{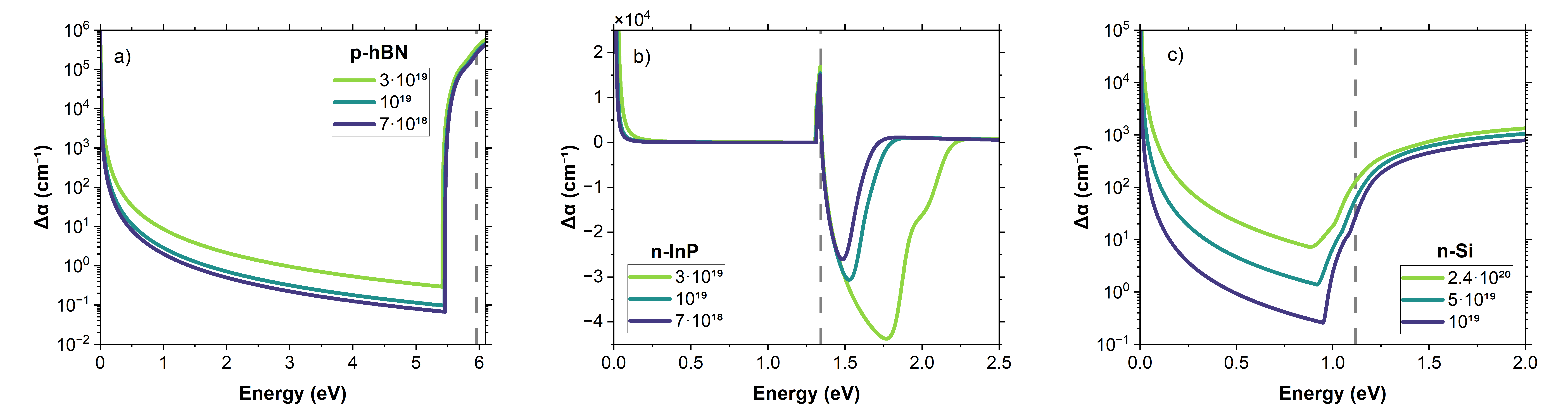}
\caption{The dependence of the change of absorption coefficient on the energy of the incident photon for (a) p-type hBN, (b) n-type doped InP and (c) n-type Si.}
\label{fig:da_all}
\end{figure}

In the first step we estimate the impact of carrier concentration on refractive index. 
Fig.~\ref{fig:da_all} presents the spectral dependence of changes in absorption coefficient due to all effects combined for hBN, InP and Si. 
Individual curves correspond to different concentrations of electrons N or holes P.
The limited concentration range was chosen based on the experimentally obtained concentrations reported in the literature, which correspond to $P = 3\cdot10^{19}$~$cm^{-3}$ for Be-doped hBN~\cite{He2009}, $N=1.1 \cdot 10^{20}$~$cm^{-3}$ for Si-doped InP~\cite{Zheng2000} and $N=10^{20}$~$cm^{-3}$ for As-doped Si~\cite{Ginn2011}.
The energy gap of intrinsic material is marked by a dashed line.
In the spectral range up to $E_g-E_{ph}-\Delta E_g(X)$, where the absorption edge occurs, the only effect that alters the absorption coefficient is the plasma effect.
Above this value, bandfilling and bandgap renormalization are major effects and final dependency is determined by their mutual relation.
Bandfilling is an effect that mainly affects the conduction band usually. It does not have a significant impact for p-type semiconductors, due to high density of states in the valence band.
Additionally, since achieving n-type conductivity in hBN seems rather difficult~\cite{Lu2022}, we decided to concentrate on the p-type one.
Moreover, hBN is material which valleys in band structure are wide and nearly flat~\cite{Cassabois2016}.
These facts contribute to the weak bandfilling effect, because of very high density of states possible to occupy at the bottom of the conduction band.
Due to the fact that bandfilling is usually stronger effect than bandgap renormalization, for InP and Si we choose to focus on n-type conductivity.
The redshift of the absorption edge is a consequence of the renormalization of the bandgap $\Delta E_g$.

\begin{figure}[htbp]
\centering\includegraphics[width=13cm]{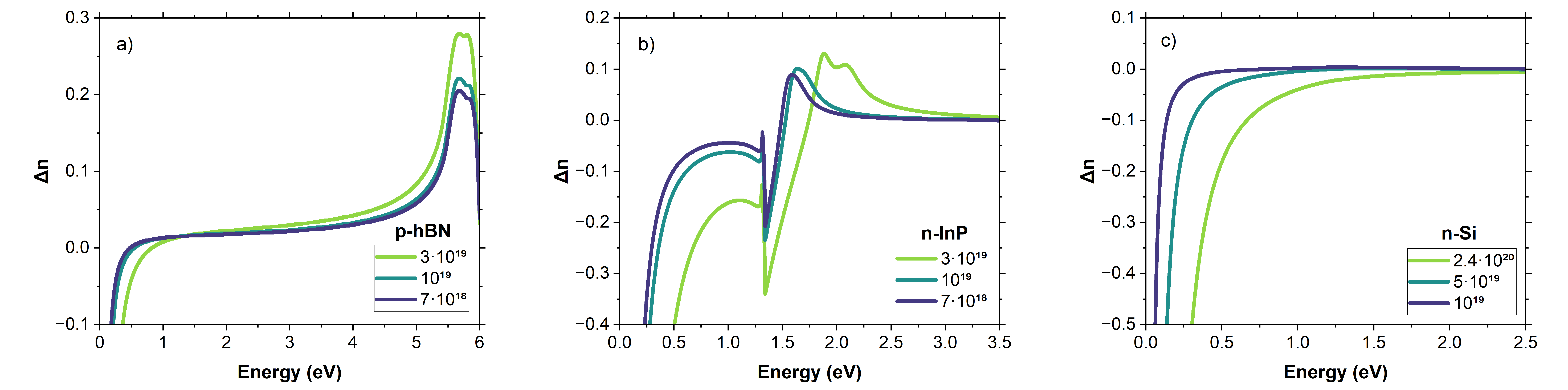}
\caption{The dependence of the change in refractive index on the energy of the incident photon for (a) p-type hBN, (b) n-type doped InP and (c) n-type Si.}
\label{fig:dn_all}
\end{figure}

\begin{figure}[htbp]
\centering\includegraphics[width=13cm]{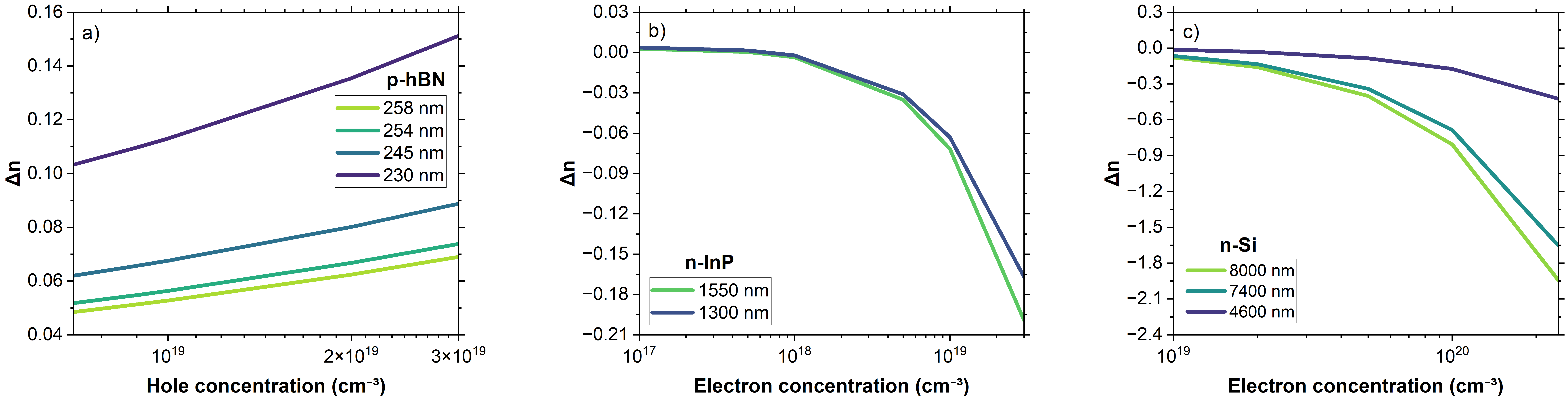}
\caption{The dependence of the change of refractive index on the free carrier concentration for (a) p-type hBN, (b) n-type doped InP and (c) n-type Si.}
\label{fig:dn_X}
\end{figure}

Fig.~\ref{fig:dn_all} presents the $\Delta n$ change with energy E for all effects combined. 
Fig.~\ref{fig:dn_X} shows the same, but as a function of carrier concentration. 
The energies at which the calculations were performed were chosen based on state-of-the-art light sources and passive elements~\cite{You2020,Benyoucef2013,Kuroda2007,Jean-Marc2018,Lim2024}, including the telecommunication windows.
In the low-energy region, absorption drastically increases, and, as a consequence, the refractive index suddenly drops.
Near the energy gap $\Delta n$ rises, which is clearly visible for p-hBN and n-InP.
Greater carrier concentrations result in higher refractive-index contrast, and therefore higher reflectivity. 
However, it also brings greater absorption of the material, which will decrease the amount of reflected light. 
Eventually, in order to achieve the best performance of the DBR, we need to adjust the number of layers.
Typically, the more layers in a DBR structure, the higher the reflectivity. 
Including free-carrier absorption results in saturation for a specific number of layers and then a drop in reflectivity.
In our further calculations, we choose the number of layers that is equal to or smaller than the amount of layers corresponding to saturation.

Reflectivity spectra calculations were performed for carrier concentrations that give the greatest contrast of refractive index and the least absorption: $P=3\cdot10^{19}$~$cm^{-3}$ for hBN, $N=3\cdot10^{19}$~$cm^{-3}$ for InP and $5\cdot10^{19}$~$cm^{-3}$ for Si.
Representative wavelengths for which DBR structures were designed are 230~nm, 1300~nm and 1550~nm, and 8000~nm, respectively.

\begin{figure}[htbp]
\centering\includegraphics[width=13cm]{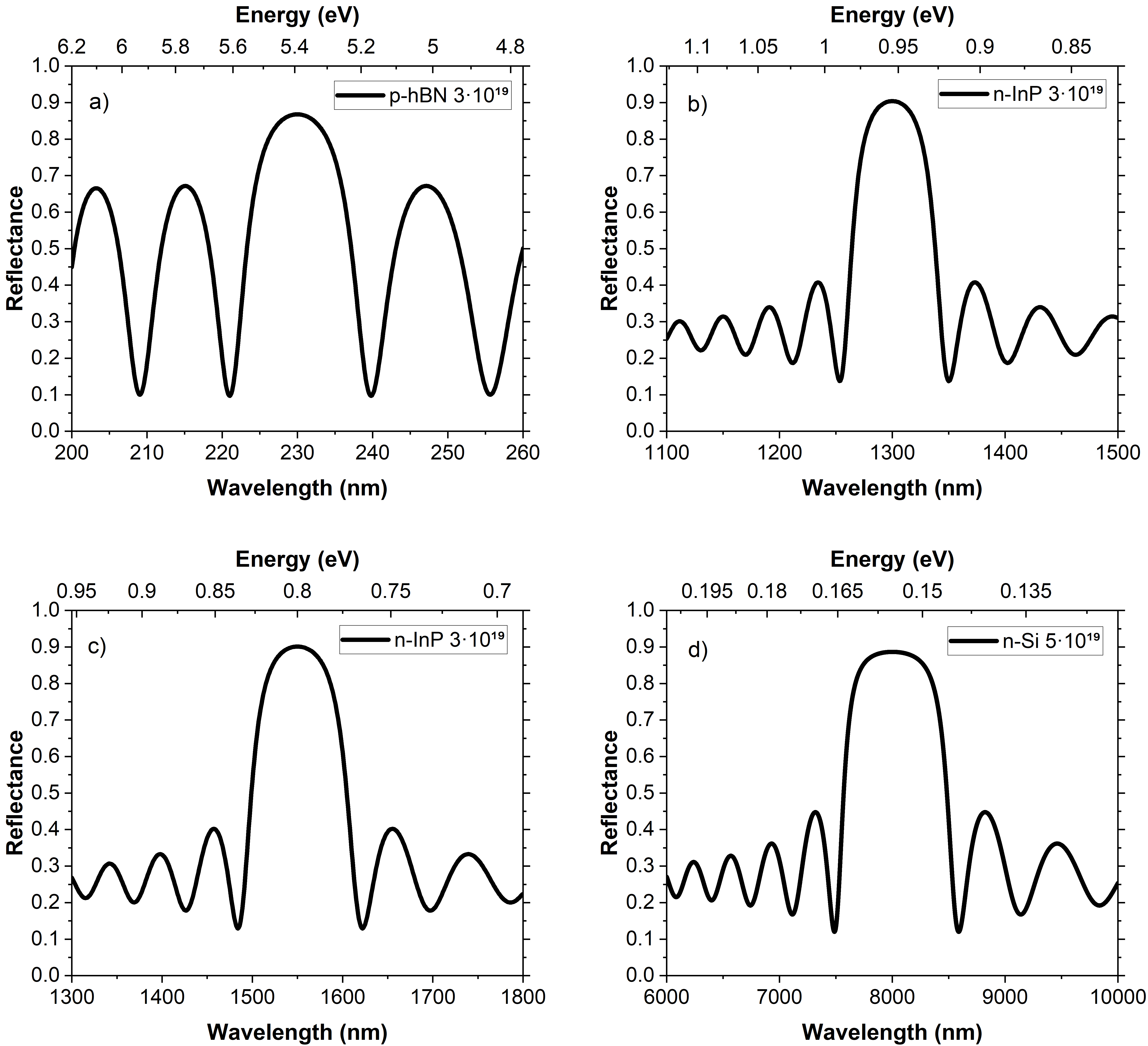}
\caption{Reflectivity spectra of DBR of (a) p-hBN for 230 nm, (b) n-InP for 1300~nm, (c) n-InP for 1550~nm, and (d) n-Si for 8000~nm.}
\label{fig:DBRs}
\end{figure}

The performance of the proposed DBR structures is shown in Fig.~\ref{fig:DBRs}, and their parameters are presented in Table~\ref{tab:parameters}.
The rows refer to different properties such as: $\lambda$ - wavelength of the of the stopband centre in $nm$ and $\Delta n$ - refractive index contrast corresponding to this $\lambda$.
Next, there are thicknesses of undoped $d_1$ and doped $d_2$ layers in the DBR structure and number of pairs of layers $m$, as illustrated in Fig.~\ref{DBR_sructure}.
Then, there are values of free carrier absorption $\Delta \alpha$ for specific $\lambda$, maximum reflectivity of structure $R_{max}$, full width at half maximum $FWHM$ and total thickness of structure $d$.

\smallskip

\begin{table}[htbp]
\centering
\caption{\bf General parameters and optical properties of proposed DBRs}
\begin{tabular}{ccccc}
\hline
Parameters & a & b & c & d \\
\hline
$material$ & hBN & InP & InP & Si \\
$\lambda (nm)$ & 230 & 1300 & 1550 & 8000 \\
$\Delta n$ & 0.149 & -0.167 & -0.199 & -0.402 \\
$\Delta n_{bf}$ & 0 & -0.0814 & -0.0701 & $-3.6659\cdot10^{-5}$ \\
$\Delta n_{br}$ & 0.1518 & 0.0124 & 0.0101 & 0.004 \\
$\Delta n_{fca}$ & $-5.6115\cdot10^{-4}$ & -0.0979 & -0.1392 & -0.405 \\
$d_1 (nm)$ & 13.02 & 101.43 & 122.44 & 576.40 \\
$d_2 (nm)$ & 4.27 & 107.00 & 130.66 & 651.93 \\
$m$ & 15.5 & 25 & 21 & 15 \\
$\Delta \alpha (cm^{-1})$ & 0.2963 & 28.2620 & 40.1901 & 48.487 \\
$R_{max} (\%)$ & 86.8 & 90.4 & 90.1 & 88.7 \\
$FWHM (nm)^{a}$ & 14.7 & 76.4 & 109.1 & 937.3 \\
$d (\mu m)$ & 0.404 & 5.243 & 5.357 & 18.906 \\
\hline
\end{tabular}
  \label{tab:parameters}\\
$^a$Stopband width at half of the maximum reflectivity.
\end{table}

\section{Discussion}
The conducted research has shown that the designed Bragg mirrors employing the refractive index contrast induced by changes in the layers' doping exclusively exhibit high reflectivity in a specific wavelength ranges of application relevance.
The difference of the refractive indices $\Delta n$ varies from 0.1 to 0.4 based on the material and the selected wavelength.
As a consequence, with 15-17 layer pairs, we are able to achieve a reflectance of 87-90\%.

hBN-based DBR has parameters comparable to the those obtained by Ciesielski et al.~\cite{Ciesielski2024} where material porousitity has been used to tailor the refractive index.
However, they structure was designed for around 700~nm to match the spectral range of optical activity of TMDCs, unlike ours for 230~nm in the UVC.
The spectral range of their choice would give in our case poor results, because $\Delta n$ nearly equals 0 and there is high absorption, as shown in Fig.~\ref{fig:dn_all}(a) and Fig.~\ref{fig:da_all}(a), respectively.

On the other hand, the InP-based DBR gave 90\% with 21-25 layer pairs. 
Badura et. al reported on DBR with 99.5\% reflectivity with 8 layer pairs. 
Better results are understandable due to the fact that the structure operates at longer wavelengths, where the difference in the refractive indices is incomparably greater and amounts to approximately $\Delta n=2.8$.
They dealt with absorption by modulating the thicknesses of layers to sustain the condition of constructive interference for the layer pair, not simply for each individual film, as considered in our research~\cite{Badura2024}.

In the case of the Si-based DBR, we achieved a reflectivity of 88.7\% using 15 pairs of alternating layers. 
In contrast, $SiO_2/TiO_2$-based mirrors demonstrate significantly higher performance, reaching reflectivity levels exceeding 95\% with only 5 layer pairs~\cite{Feng2013}. 
These findings indicate that our structure exhibits comparatively lower optical performance, although might be advantagous technologically.

Better results, that is, greater reflectivity, may be obtained by adding more layer pairs, with the limit determined by absorption, or by increasing the refractive index contrast through carrier concentration modulation.
With approximately 40 layer pairs, a reflectivity of around 98\% can be achieved in the range of low absorption.
The maximum reflectance we can achieve is 99.8\%, but the number of layer pairs is striving towards a hundred.

The main advantage of the proposed DBR structures is the possibility of a homoepitaxial growth process, which simplifies fabrication by utilizing a single material.
This uniformity also eliminates issues related to strain accommodation caused by lattice constant mismatches, as induced stress can lead to structural damage.

Doped layers offer an additional advantage — they are inherently suitable for injection of electrical carriers into the active region. 
This makes them particularly promising for achieving current flow in DBR-based optoelectronic devices. 
However, for this concept to be viable, the material must also exhibit high carrier mobility. 
Among the proposed materials, only InP and Si possess a sufficiently high carrier mobility to support conduction, making them the most suitable candidates for such an application. 

However, it should be noted that the reflectivities achieved in our structures may not be sufficient for highly demanding applications such as VCSELs, which require extremely high feedback. 
Nevertheless, they can effectively enhance emission from single-photon sources or other quantum emitters, where moderate reflectivity is adequate to improve the performance.

\section{Conclusions}
In our study we designed and theoretically investigated optical properties of homoepitaxial DBRs where the refractive index contrast is induced by increasing the carrier concentration into every second layer, e.g., by doping. 
This approach lifts up-to-date restriction of material choice for the two layers forming the DBR to be lattice-matched, which paves the way towards new DBR designs for materials and spectral ranges of so far poor performance.  
The calculations considered three different semiconductor material systems, relevant for selected optoelectronic and nanophotonic applications: hBN, InP, and Si.
Our approach resulted in approximately 90\% reflectivity with 15-25 layer pair, depending on the material.
Such structures could serve as mirrors to enhance extraction efficiency of photons from compatible emitters~\cite{You2020,Benyoucef2013,Kuroda2007,Jean-Marc2018,Lim2024} or to create cavities to tailor the light-matter interaction and to increase the photon generation rates.
However, precise control of growth process and developed doping techniques are necessary, in particular localization of dopants only in one of the two layers constituting the DBR.
These results can serve as a base for further developments of homoepotaxial DBRs and we hope they will inspire the technologists. 
They can also contribute to the development of homoepitaxial, fully conducting distributed Bragg reflectors for optoelectronic applications utilizing carrier injection into the active region. 
In the case of advanced nanophotonic applications, they enable electrical control of the charge state of the quantum emitter and charge fluctuations of its environment. 
This is crucial for long coherence times of the carrier charge or spin state and the indistinguishability of emitted photons, which is particularly important for the preservation of the qubit state, realization of quantum memories and for quantum repeater architectures.

\begin{backmatter}
\bmsection{Acknowledgment}
The project was funded by the QuanterERA II European Union’s Horizon 2020 research and innovation programme under the EQUAISE project, Grant Agreement No. 101017733, under the supervision of the National Center for Research and Development in Poland within the project QuantERA II ERA-Net Cofund in Quantum Technologies (QUANTERAII/1/74/EQUAISE/2022). 

This work was co-financed by the Ministry of Education and Science, Republic of Poland within the „Perły Nauki” project, grant no. PN/01/0117/2022.

\bmsection{Disclosures}
The authors declare no conflicts of interest.

\bmsection{Data Availability Statement}
Data underlying the results presented in this paper are not publicly available at this time but may be obtained from the authors upon reasonable request.
\end{backmatter}
\bibliography{sample}

\end{document}